\documentstyle[eqsecnum,epsfig,twocolumn,aps,floats,graphicx]{revtex}

\begin{document}

\twocolumn[\hsize\textwidth\columnwidth\hsize\csname
@twocolumnfalse\endcsname
\title{Momentum-Resolved Charge Excitations in a One Dimensional
Prototype Mott Insulator}

\author{M. Z. Hasan$^{1}$, P. A. Montano$^{2}$, E. D. Isaacs$^{3}$, Z.-X. Shen$^{1}$,
H. Eisaki$^{1}$, \linebreak S. K. Sinha$^{4}$, Z. Islam$^{4}$, N.
Motoyama$^{5}$, S. Uchida$^{5}$}

\address{$^1$Department of
Applied Physics and Stanford Synchrotron Radiation Laboratory
(SSRL) of Stanford Linear Accelerator Center(SLAC), Stanford
University, Stanford, CA 94305}
\address{$^2$Department of Physics, University of Illinois at Chicago, Chicago, IL 60607}
\address{$^3$Bell Laboratories, Lucent Technologies, Murray Hill, NJ 07974}
\address{$^4$Advanced Photon Source, Argonne National Lab, Argonne, IL 60439}
\address{$^5$Department of Superconductivity, University of Tokyo, Bunkyo-ku, Tokyo 113, Japan}

\date{\today}
\maketitle


\begin{abstract}

We report momentum resolved charge excitations in a one
dimensional (1-D) Mott insulator studied using high resolution
($\sim$ 325 meV) inelastic x-ray scattering over the entire
Brillouin zone for the first time. Excitations at the insulating
gap edge are found to be highly dispersive (momentum
dependent)compared to excitations observed in two dimensional Mott
insulators. The observed dispersion in 1-D is consistent with
charge excitations involving holons which is unique to spin-1/2
quantum chain systems. These results point to the potential
utility of inelastic x-ray scattering in providing valuable
information about electronic structure of strongly correlated
insulators.

\end{abstract}

\vspace*{0.21 in}

]

\narrowtext After several decades of research efforts, electronic
structure of late transition metal oxides lacks comprehensive
understanding. The existence of exotic electronic, magnetic and
optical properties such as high T$_c$ superconductivity as
exhibited by the cuprates or colossal magnetoresistance as in the
manganites or highly nonlinear optical responses as observed in
the nickelates are believed to be related to the strong
electron-electron Coulomb correlations in these systems [1-4].
This suggests the necessity of studying their  correlated charge
dynamics. In last several years, with the advent of high
brightness synchrotron facilities, inelastic x-ray scattering has
been developing as a tool to study the BULK electronic structure
of condensed matter systems [4-10]. X-ray scattering from the
valence charge distribution is fairly weak thus difficult to
distinguish from the total scattering signal especially in high-Z
materials making such experiments quite difficult to perform.
Recent experimental and theoretical investigations have shown that
by tuning the incident energy near an x-ray absorption edge a
large enhancement can be achieved making the study of valence
excitations feasible in high-Z systems [5,7-10]. One dimensional
half-filled spin-1/2 quantum systems as realized in some copper
oxide compounds (such as Sr$_2$CuO$_3$ and SrCuO$_2$) believed to
exhibit spin-charge separation. As a consequence, in these
systems, charge fluctuations propagate rather freely and
independently of the spin fluctuations [11-15]. This is in
contrast to the two dimensional (2-D) spin-1/2 systems where
charge motion is strongly coupled to the spin fluctuations and
rather restricted [10,12,14-17]. In 1-D, charge excitations would
be expected to be highly dispersive compared to analogous 2-D
systems. Because of the insulating gap, charge fluctuations are at
relatively high energies in Mott insulators [1,2]. The momentum
dependence of the effective Mott gap (charge-transfer gap) has
been reported recently in a parent 2-D cuprate using inelastic
x-ray scattering [8,10,20]. \linebreak In this Letter, we report
study of 1-D Mott insulator's charge fluctuation spectrum by
varying \textbf{q} (the scattering vector) over the entire
Brillouin zone using high resolution inelastic x-ray scattering.
The Mott gap is found to be of a direct nature (minimum of the gap
appears at \textbf{q} $\sim$ 0) within the level of experimental
resolution and the charge fluctuations at the gap edge are more
dispersive along the Cu-O bond direction in 1-D as compared to 2-D
parent cuprate insulator.

The experiment was performed using the high flux undulator
beamline 12-ID (BESSRC-CAT) at the Advanced Photon Source of
Argonne National Laboratory. Inelastic scattering was measured by
varying \textbf{q} along the chain direction (Cu-O bond direction)
of single crystalline Sr$_2$CuO$_3$. Overall energy resolution of
325 meV was achieved for this experiment. This is an improvement
over our earlier works on 2-D Mott systems by more than 100 meV
[8,10,20]. This improvement in resolution (in combination with the
high flux from the Advanced Photon Source) allowed us to resolve
the Mott excitations in 1-D systems for the first time despite
high-level of x-ray absorption due to Sr in the system under
study. The energy of the incident beam was set near the Cu K-edge
(E$_o$ = 8.996 eV) for resonant enhancement of excitation
features. The scattered beam was reflected from a diced Ge-based
analyzer and focused onto a solid-state detector. For
\textbf{q}-scans, the incident energy was kept fixed and
\textbf{q} was varied by rotating the entire spectrometer around
the scattering center. The background, measured on the energy gain
side, was about 2-3 counts per minute. Sr$_2$CuO$_3$ crystals used
for this experiment were grown and characterized by techniques
described previously which confirmed its quasi-one dimensionality
above 6 K (N\'{e}el transition due to 3-D coupling)[14,15]. Unlike
extensively studied 1-D cuprates such as CuGeO$_3$ or KCuF$_3$,
Sr$_2$CuO$_3$ and SrCuO$_2$ show no spin-Peierls transition hence
provides a unique opportunity to study the charge fluctuations in
a 1-D spin-1/2 quantum Heisenberg chain system [14,15].

\begin{figure}[t!]
\centerline{\epsfig{figure=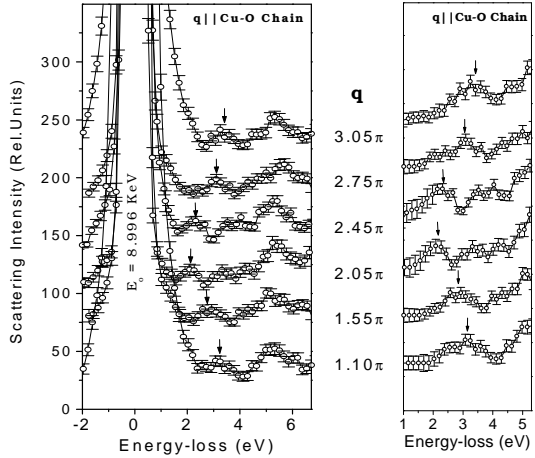,width=8.7cm,clip=}}
\vspace{.3cm}  \caption{ Inelastic x-ray scattering spectra along
the chain direction (the Cu-O bond direction) are shown as a
function of q scanning over the entire Brillouin zone. The values
of q for the spectra bottom to top are 1.10$\pi$, 1.55$\pi$,
2.05$\pi$, 2.45$\pi$, 2.75$\pi$ and 3.05$\pi$ respectively. The
spectrum for q $\sim$ 1.1$\pi$ was taken with a resolution of
about 490 meV whereas the rest were taken with 325 meV energy
resolution. The right panel focuses on the lower energy feature
(elastic scattering removed by fitting). E$_o$ = 8.996 KeV}
\label{fig1}
\end{figure}

Fig.1 shows inelastic x-ray scattering spectra with varying
momentum transfers along the chain direction (the Cu-O bond
direction) with incident energy fixed near Cu K-edge (E$_o$ =
8.996 KeV). All the spectra in each panel were normalized to the
intensity in a window between 8 and 9 eV energy-loss (not shown in
the Fig.1). Each spectrum shows two features, one around 5.6 eV
and another, lower in energy, appear in the range of 2.5 to 3.5 eV
depending on different values of the scattering wave vector,
\textbf{q}. The 5.6-eV feature can be assigned to be a charge
transfer excitation from the groundstate to the antibonding-type
excited states which is analogous to the 6 eV excitation observed
in 2-D cuprate insulators [7,8,10]. The other prominent feature
that appears at lower energies has a significant movement in
changing \textbf{q}. The feature disperses upward in energy about
1 eV monotonically over the Brillouin zone in going from the zone
center (q $\sim$ 2$\pi$) to the edge of the zone (q $\sim$ $\pi$,
$3\pi$). Inelastic features with similar intensities and
dispersions were also seen for incident energy of 8.992 KeV.

\begin{figure}[b!]
\centerline{\epsfig{figure=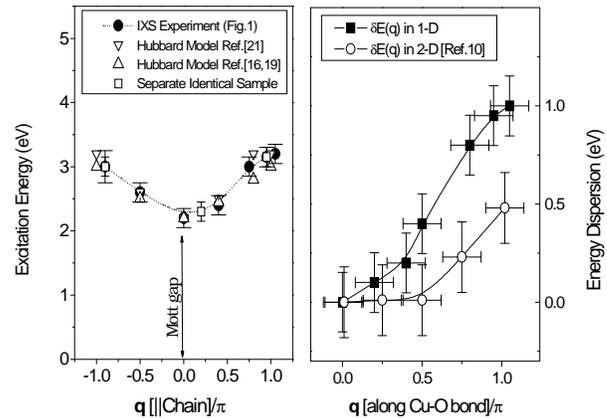,width=8.7cm,clip=}}
\vspace{.2cm} \caption{(Left panel) Momentum dependence of the
experimental inelastic features are compared with different types
of numerical calculations [21,16,19] based on the Hubbard model.
Data points with square symbols are from an independent
experimental run not shown in Fig-1. (Right panel) Comparison of
\textbf{q}-dependence of charge excitations along the Cu-O bond
direction in 1-D and 2-D [10] is shown. All the q points are
plotted in their equivalent positions within 0 to $\pi$ for
comparison. Charge excitations are found to be more dispersive in
1-D than 2-D.} \label{fig2}
\end{figure}
Inelastic x-ray scattering measures the dynamical charge-charge
correlation function (charge fluctuations) which can be
interpreted as particle-hole pair excitations in the range of
momentum-transfers comparable to the size of the Brillouin zone of
the system. Near an absorption edge the measured response function
gets modified but it can still be interpreted as composites of
pair excitations [8-10,18-21]. In a simplistic view, the core-hole
created by the X-ray photon near the absorption edge causes
electronic excitations in the valence band which can be composed
of having a hole in the occupied band and an  electron in the
unoccupied band across the gap. We interpret the  dispersion of
the low energy feature seen in our data as the dispersion of the
effective Mott gap (charge-transfer type [22]) in the system.  The
particle-hole pair formed in the process absorbs the energy lost
from the incident x-ray beam and propagates with momentum
$\hbar$\textbf{q}. The propagation of this pair would depend on
the charge and spin distributions in the system. Sr$_2$CuO$_3$ and
SrCuO$_2$ are believed to be quasi-1-D half-filled Mott insulators
with short-range antiferromagnetic spin correlations (quantum
Heisenberg systems) [14,15] and has a Cu-O bond length (lattice
constant of 3.90 $\AA$) much similar (less than 1\% difference) to
the Cu-O bond length of 2-D parent cuprate such as
Ca$_2$CuO$_2$Cl$_2$ (lattice constant of 3.87 $\AA$) studied
earlier [10,14].

Spectroscopic studies interpreted based on model calculations
suggest that these 1-D cuprates exhibit spin-charge separation
[14-16]. Angle-Resolved Photoemission Spectroscopy (ARPES) shows
that the hole bandwidth is much larger in 1-D than 2-D contrary to
the expectation of LDA-type model for electronic structure [15].
The quasiparticle decay modes , probed by ARPES, are interpreted
as spinons and holons [15]. Since holon is a collective charge
mode it would couple to the x-rays strongly and exhibit its
characteristic \textbf{q}-dependence. It is interesting to note
that \textbf{q}-dependence of the Mott feature in 1-D is larger
than it is along the bond direction in 2-D we studied earlier
[10]. We compare this behavior in Fig. 2(Right panel). Such
behavior would be qualitatively expected when charge fluctuations
are free to move because of decoupling from the spin degrees of
freedom. This is also seen from numerical studies of Hubbard model
[19,21]. In Fig. 2(Left panel) we compare our experimental results
to the expectations from 1-D half-filled spin-1/2 Hubbard model
describing charge fluctuation spectrum at finite-q [16,19,21].
Within the level of energy resolutions the results are
qualitatively described by (or at least consistent with)
excitations involving holons in 1-D Hubbard model. It is
interesting to note that these results are qualitatively
consistent with electron scattering (nonresonant) studies of 1-D
cuprates mainly in the low-energy regime [16].

\textbf{q}-dependent charge fluctuations in 1-D Mott insulators
indicate that in 1-D Mott gap is of direct nature and excitations
at the gap edge are more dispersive as compared to 2-D. The
dispersions are also consistent with models describing the motion
of holons in 1-D spin-1/2 Mott insulators. These results suggest
that inelastic x-ray scattering can be used to study electronic
structure of complex insulators and correlated electron systems in
general. Higher resolution experiments would be necessary to
extract quantitative details about the fundamental electronic
parameters using such spectroscopies. Development of high
brightness fourth generation synchrotrons can potentially make
such experiments feasible with energy resolution in the millivolt
regime.

The experiments were performed at the Advanced Photon Source (APS)
of Argonne National Laboratory. APS is supported by the U.S. DOE,
Office of Science, under Contract No. W-31-109-ENG-38. This work
was also jointly supported by the Department of Energy through
Stanford Synchrotron Radiation Lab of Stanford Linear Accelerator
Center, Stanford, California and Bell-Laboratories of Lucent
Technologies, New Jersey.

\vspace*{-0.2in}

\begin{references}
\bibitem{hubbard}  J. Hubbard  Proc. Phys. Soc. London A  \textbf{277}, 237 (1964).
\bibitem{mt} N. F. Mott Metal Insulator Transitions, Taylor
Francis Ltd., London (1974).
\bibitem{anderson} P. W. Anderson, Science \textbf{235}, 1996
(1987).
\bibitem{tokura} Y. Tokura and N. Nagaosa Science \textbf{288}, 462
(2000).
\bibitem{kao} C. C. Kao, W. A. L. Caliebe, J. B. Hastings, J. M. Gillet  Phys. Rev. B \textbf{54},
16 361 (1996).
\bibitem{isaacs}  E. D. Isaacs, P. M. Platzman, P. Metcalf, J. M. Honig Phys. Rev.
Lett. \textbf{76}, 4211 (1996).
\bibitem{jhill} J. P. Hill {\sl et al.}, Phys. Rev. Lett. \textbf{80}, 4967 (1998).
\bibitem{abbamonte} P. Abbamonte {\sl et al.}, Phys. Rev. Lett. \textbf{83}, 860 (1999).
\bibitem{platzman}P. M. Platzman  and E.D. Isaacs Phys. Rev. B \textbf{57}, 11 107
(1998).
\bibitem{hasan}M. Z. Hasan {\sl et al.}, Science \textbf{288}, 1811
(2000).
\bibitem{PWA}P. W. Anderson, A Career in Theoretical Physics, World Scientific
Series in 20th Century Physics (World Scientific,
Singapore,(1994), Vol. \textbf{7}.
\bibitem{laughlin}R. B. Laughlin  Phys.  Rev. Lett. \textbf{79}, 1726 (1997).
\bibitem{lieb}E. H. Lieb and F. Y. Wu   Phys. Rev. Lett. \textbf{20}, 1445 (1968).
\bibitem{motoyama}N. Motoyama, H. Eisaki and S. Uchida   Phys. Rev. Lett. \textbf{76},
3212 (1996).
\bibitem{kim}C. Kim {\sl et al.}, Phys. Rev. Lett.  77, 4054 (1996); C. Kim et.al., Phys. Rev. B \textbf{56}, 15589 (1997).
\bibitem{neudert}  R. Neudert {\sl et al.}, Phys. Rev. Lett. \textbf{81}, 657 (1998).
\bibitem{bwells}  B. O. Wells {\sl et al.}, Phys. Rev. Lett. \textbf{74}, 964 (1995).
\bibitem{tsutsui} K. Tsutsui, T. Tohyama, S. Maekawa Phys. Rev. Lett. \textbf{83}, 3705 (1999).
\bibitem{stephan}   W. Stephan and  K. Penc Phys. Rev. B \textbf{54}, R 17269 (1996).
\bibitem{zhasan}  M. Z. Hasan {\sl et al.}, Physica C \textbf{341-348}, 781
(2000).
\bibitem{ktsutsui}  K. Tsutsui, T. Tohyama, S. Maekawa Phys. Rev. B \textbf{83}, 3705 (1999).
\bibitem{zsa}   J. Zaanen, G. A. Sawatzky, J. W. Allen  Phys. Rev. Lett. \textbf{55}, 418 (1985).


\end{references}
\end{document}